# Advanced Modeling for Exoplanet Detection and Characterization

Krishna Chamarthy
Department of Computer Engineering and Technology
Dr. Vishwanath Karad MIT World Peace University,
Pune, India
1032221617@mitwpu.edu.in

*Abstract*— Research into light curves from stars (temporal variation of brightness) has completely changed how exoplanets are discovered or characterised. This study including star light curves from the Kepler dataset as a way to discover exoplanets (planetary transits) and derive some estimate of their physical characteristics by the light curve and machine learning methods. The dataset consists of measured flux (recordings) for many individual stars and we will examine the light curve of each star and look for periodic dips in brightness due to an astronomical body making a transit. We will apply variables derived from an established method for deriving measurements from light curve data to derive key parameters related to the planet we observed during the transit, such as distance to the host star, orbital period, radius. The orbital period will typically be measured based on the time between transit of the subsequent timelines and the radius will be measured based on the depth of transit. The density of the star and planet can also be estimated from the transit event, as well as very limited information on the albedo (reflectivity) and atmosphere of the planet based on transmission spectroscopy and/or the analysis of phase curve for levels of flux. In addition to these methods, we will employ some machine learning classification of the stars (i.e. likely have an exoplanet or likely do not have an exoplanet) based on flux change. This could help fulfil both the process of looking for exoplanets more efficient as well as providing important parameters for the planet. This will provide a much quicker means of searching the vast astronomical datasets for the likelihood of exoplanets.

## I. INTRODUCTION

Over the past three or four decades, the fascinating field of astrophysics was concerned with studying the presence of exoplanets, which are planets orbiting stars external to our solar system. In thousands of such worlds, scientists found planets of a variety of sizes and orbits. When an exoplanet transits a star, which is measuring how the star dims at regular periodic intervals, this is one of the main methods used for finding exoplanets. It was the photometry and spectrometry techniques that did both-the detection of exoplanets along with giving information about their diameters, orbits, and atmospheric compositions.

Stellar light curve analysis, in which the star's brightness or flux over time is monitored, forms the basis for transit-based detection. The light curves are highly informative, and the astronomers can establish if the periodic variation in brightness is due to planetary transits or some other astrophysical activity. From the observation of the variation in light depth, shape, and timing, the most crucial exoplanetary parameters, like size, orbital period, and host-star distance, are capable of being estimated. The observation also assists in more advanced techniques like transmission spectroscopy in measuring the composition of the planet's atmosphere by examining the manner in which starlight of different wavelengths is absorbed during transit.

Here, we analyze stellar light curves from the Kepler dataset, comprising observations of nearly 2,00,000 stars collected by the Kepler space telescope over a couple of years, to discern exoplanets and draw inferences about pertinent planetary parameters. With the help of machine learning models trained on flux data, we attempt to automate the detection of exoplanet candidates and classify stars as non-planet-hosting or planet-hosting. In addition to rudimentary detection, we draw inferences about important planetary properties, i.e., radius and orbital period, based on transit-based relations. The potential for detecting and characterizing exoplanets based on light curves holds out the promise of scaling the exoplanet search in large data sets, allowing for the detection of potentially habitable planets in star systems light-years away. We employ Random Forests for data preprocessing because of their capability to detect and handle anomalies and outliers in the flux measurements. This operation provides a cleaner dataset by detecting problematic data points, thus enhancing the quality of the following analysis. Complicated patterns in the light curves are investigated using Convolutional Neural Networks (CNNs) to have a better insight. CNNs possess the capability to boost the detection of faint transit signals and facilitate the extraction of detailed planetary attributes by learning hierarchical features from raw data automatically, enhance the detection of faint transit signals and facilitate the extraction of detailed planetary attributes.

## II. LITERATURE REVIEW

The AI methodology has given a revolutionary change in the detection and characterization of exoplanets. The domain has now shifted from conventional observation techniques to AI methods that provide more precision and efficiency in exoplanet detections.

The research to date confirmed the pioneering role of AI in exoplanet discovery. According to Liu et al. (2024) [1], machine learning algorithms have taken centre stage when working with large datasets from the likes of Kepler, TESS, and so on. These algorithms detect exoplanet candidates by automatically identifying periodic dips in a star's brightness, i.e., transits. The use of convolutional neural networks (CNNs) has, in particular, been beneficial for detecting subtle patterns from light curves for exoplanet marking with an accuracy and scale that would have been virtually impossible otherwise by manual means.

The work of Ricker et al. (2021) [2] proves the usefulness of Random Forest algorithms in the pre-processing and cleaning of astronomical data. They are especially helpful to work against the noise and inconsistency in flux measurements, guaranteeing data quality and furthering the reliability of results obtained downstream in the analysis. This pre-processing is very important because it ensures that machine learning models are trained on clean, valid data sets required to identify potential exoplanets correctly. First, classifiers for both machine learning on the Kepler dataset presented an accuracy of 94%. Following this, to improve this accuracy further, scientists coupled the system with deep learning approaches, and the accuracy later soared to 98.9% on the TESS data set, proving an instance of the hybrid strength of conventional machine learning with deep learning approaches in exoplanet detection.

Accuracy maintenance of the classification model is achieved by data quality management. Our approach utilizes Random Forest algorithms for outlier identification and correction in the preprocessing stage, thus greatly enhancing the quality of the dataset before the modelling stage. It is a crucial preparation step toward enhancing the working capability of any future machine learning model.

The technique uses a blend of both traditional machine learning and the state-of-the-art deep learning methods. Specifically, Convolutional Neural Networks (CNN) are used to learn those high-level hierarchical features from the raw light curve data that are more suited for analysis of complex time-series astronomical signals. Additionally, the RNN implementation is intended to recognize sequential patterns and build temporal relationships for the enhancement of exoplanet detection.

Having combined these machine learning and deep learning techniques can also boost detection power and precision of characterization; better light curve data allows us to apply the transit equations and obtain important planetary parameters such as radius measurements, orbital period estimation, and stellar distance computations. This hybrid detection and characterization method stands far more advanced for exoplanet research, thereby proving an incredible landmark of AI on the current astrophysics' scene. With this integration between machine learning and deep-learning methods, classification of exoplanets becomes easy and refined, thus opening the gateway for automating astronomical discovery. As AI advances, future research will build upon these methods and develop techniques to integrate AI in present data sets from observations. These improvements will garner the ability in understanding planetary systems and searching for potential habitats.

## III. PROPOSED METHOD

Our exoplanet-detection method is a hybrid deep learning-traditional machine learning system that combines convolutional neural networks (CNN) with Random Forest classifiers. Our multi-layer CNN architecture processes sequential stellar light curve data to automatically learn features and find faint temporal patterns signifying exoplanetary transits. The process extracts faint brightness variations efficiently, typical of a planet transiting its parent star. Meanwhile, the Random Forest classifier enables a strong decision-making system based on an ensemble of decision trees, which help where the observation data makes it harder to classify between exoplanet presence and absence. Any prep work on the data is integral to the system, and among the employed techniques are normalization and oversampling to combat the intrinsic class imbalance in exoplanet data. These preprocessing steps greatly embed stability into model performance. CNN's ability to handle time-series data and Random Forest's complementary classification form a potent hybrid able to capitalize on the advantages of both deep and shallow machine learning techniques. Given this hybrid, it logically follows that it will attain good levels of accuracy in exoplanet detection from stellar light curves, accurately capturing the dense patterns that signal planetary transits in astronomical recordings.

*A. Data Set Description*

The exoplanet discovery data set employed in this study contains observation data capturing stellar brightness changes, which are the signs of possible planetary transits. Derived from space missions such as NASA's Kepler or TESS, it consists of features derived from light curves and time series observations, such as flux intensity and periodogram peaks, that contain temporal patterns to detect exoplanets. The data set is classified as binary classification, in which each record is classified as showing the existence (1) or nonexistence (0) of an exoplanet. During training data preparation in a model, preprocessing such as normalization is performed so that there will be consistency in the feature scales to facilitate learning by machine learning algorithms effectively.

For handling class imbalance in the data set used in detecting exoplanets, we utilized the Random Over Sampling (ROS) method for synthesizing data. This approach produces more samples of the minority class (presence of exoplanet) by replicating existing ones randomly, thus balancing the classes. Through augmentation of training data with these synthetic samples, the model is exposed to more diversity of instances belonging to the exoplanet class and can hence learn the hidden patterns better and minimize bias toward the majority class. This method enhances the model's sensitivity to discover exoplanets, especially on datasets with limited positive instances.

## B. Feature Extraction

Feature extraction is an important part of exoplanet detection in this study because we are assessing stellar flux time series. The flux measurements indicate periodic dimming, which could mean a planet passed in front of its star at that time. The original flux measurements are generally not features meaningful to machine learning models including Random Forest and Logistic Regression. By extracting features from the original flux measurements, the models should be able to distinguish signals corresponding to true exoplanets from background signals with more precision and accuracy. The dimensionality of the flux time-series measurements can be visualized and better understood via light curve plot graphical displays of the flux changes. When used in conjunction with advanced feature extraction processes, these representations of numerical and graphical information should greatly increase the accuracy and reliability of models predicting exoplanet discoveries.

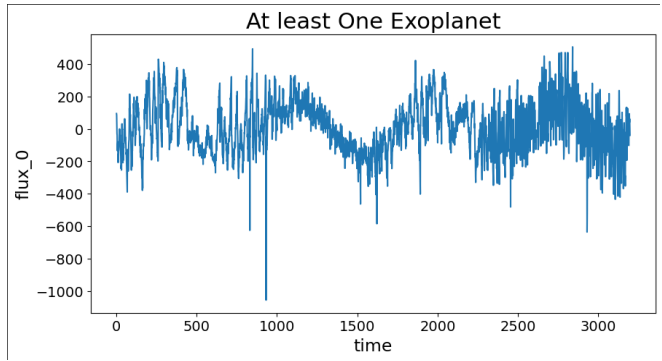

Fig. 1. Light curves graph with atleast one exoplanet

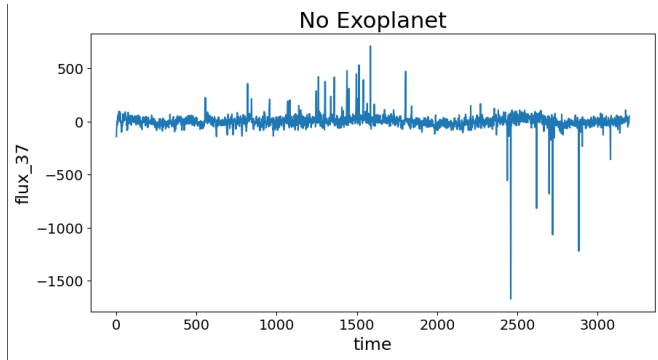

Fig. 2. Light curves graph with no exoplanets

Both graphed plots capture the fluctuations of light flux coming from stars, which are significant in finding exoplanets. The second (Fig 2.) is what usually illustrates a non-exoplanet star flux data through time, where its light is reasonably constant and just slightly affected by noise or intrinsic stellar activity. The first one (Fig 1.), as opposed to it, depicts an exoplanet star flux whose light reveals telltale regular dips. Dips are produced by these when a transit of the exoplanet across its star occurs, momentarily eclipsing the starlight to some extent. By comparing all these graphs, the obvious variance in flux curves indicates the exoplanet is present.

## C. Planetary Attributes Calculation

Apart from the roles and benefits of ML and DL models in exoplanet detection, worth mentioning is the process of determining planetary characteristics from the flux data, which provides a further layer of understanding to the exoplanet detection system. The size of an exoplanet can be determined through the transit depth, which is the contrast between the mean flux and the minimum flux that occurs during a transit. Transit depth (ΔF) is connected to the radius of an exoplanet (Rp) by the following equation:

$$\Delta F = \left(\frac{R_p}{R_*}\right)^2 \qquad (1)$$

The orbital period (P) can be estimated from time intervals in successive transits. By measuring the time intervals between the regular dips in the light curve, the time that the exoplanet is orbiting its host can be obtained. The orbital period gives us an idea of the distance that an exoplanet is from its host star, and helps in identifying if it is within the habitable zone.

The velocity of the exoplanet moving through its orbit can be estimated from its semi-major axis of orbit and orbital period. The velocity (v) can be estimated using the Keplerian period and assuming a circular orbit as:by:

$$v = \frac{2\pi a}{P} \qquad (2)$$

The light curves' periodicity and shape can offer information on orbital eccentricity, orientation, and other kinematic characteristics of the exoplanet.

## IV. DETECTION MODEL

Both the machine learning (ML) and deep learning (DL) models were used in the present study for exoplanet detection from stellar flux data with their respective advantages of dealing with the data complexity.

Among t The Random Forest Classifier algorithm is especially notable among the machine learning algorithms. It is an ensemble-based algorithm that has a collective goal of identifying nonlinear patterns and correlations that arise between high-dimensional flux data. The Random Forest Classifier was a robust algorithm, in particular, because it was able to reduce overfitting through averaging over less generalizable noise, making it effective and robust even for noisy measurements for light curves. Logistic Regression is another method that was used as a baseline algorithm, as it was easy and interpretable, and its probabilistic outputs made it easy to interpret, and understand the importance of the different flux features, and the impact they had on the classification of exoplanets. K-Nearest Neighbours (KNN) is also an easy and intuitive baseline algorithm that classifies the stars using majority voting from the close's neighbours, and therefore informing about the spatial aspects of local data. LightGBM is a used because it is a gradient-boosting library to handle large datasets with many data features and high-dimensional interactions. LightGBM is also one of the best algorithms because of its efficiency, accuracy, and because it reduces resource usage from one dataset to another, which was important when building maximum quality models when feature engineering high-dimensional datasets commonly encountered for exoplanet detections.

```
Classification Report is:
              precision    recall  f1-score   support

           0       0.99      0.99      0.99       565
           1       0.00      0.00      0.00         5

    accuracy                           0.99       570
   macro avg       0.50      0.50      0.50       570
weighted avg       0.98      0.99      0.98       570
```

Fig. 3. Report of Random Forest Model

```
Classification Report is:
              precision    recall  f1-score   support

           0       0.99      0.51      0.68       565
           1       0.01      0.60      0.02         5

    accuracy                           0.51       570
   macro avg       0.50      0.56      0.35       570
weighted avg       0.98      0.51      0.67       570
```

Fig. 4. Report of Logistic Regression Model

```
Classification Report is:
              precision    recall  f1-score   support

           0       0.99      1.00      1.00       565
           1       0.00      0.00      0.00         5

    accuracy                           0.99       570
   macro avg       0.50      0.50      0.50       570
weighted avg       0.98      0.99      0.99       570
```

Fig. 5. Report of K-Nearest Neighbours Model

```
Classification Report is:
              precision    recall  f1-score   support

           0       0.99      1.00      1.00       565
           1       0.00      0.00      0.00         5

    accuracy                           0.99       570
   macro avg       0.50      0.50      0.50       570
weighted avg       0.98      0.99      0.99       570
```

Fig. 6. Report of LightGBM Model

Deep learning algorithms like CNNs are excellent at detecting complex patterns in time-series flux measurements without human intervention in the feature engineering process. CNNs automatically learn features by recognizing dips and variations in curves caused by exoplanet transits as well as by discerning complex temporal patterns which may not be apparent to other human-driven machine learning algorithms. The deep learning methods are able to analyse data and detect subtle patterns, which only makes these methods utmost suitable for uncovering exoplanet signals confounded with noise. They learn hierarchical representations that capture both low-level features such as minuscule variations in flux and high-level abstractions such as periodic patterns better than conventional algorithms.

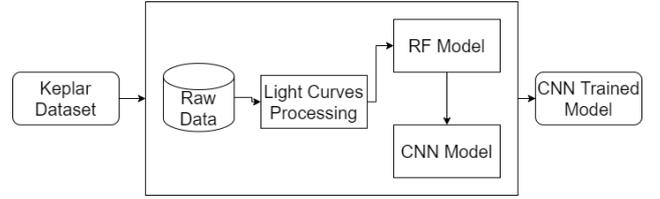

Fig. 7. Model Architecture

Through the combination of both ML and DL models, this work leverages the complementary advantages of both approaches. ML models like Random Forest and LightGBM excel in handling structured data and generating interpretable results and are thus convenient whenever interpretability and efficiency are the concern. DL models, in the form of CNNs, excel at automatic feature learning and detecting intricate patterns in unstructured time-series data. Combining both models ensures a more robust exoplanet detection system, allowing the work to balance interpretability, precision, and computational cost, and effectively handling noise and signal complexity in stellar flux data.

## V. EVALUATION METRICS

In this study, the models were evaluated using several performance metrics commonly applied in classification tasks. These metrics include accuracy, precision, recall, F1 score, and the confusion matrix. The formulas for these evaluation methods are as follows:

1) **Accuracy**: The fraction of total predictions (both transit and non-transit) that are correct

$$\text{Accuracy} = \frac{TP + TN}{TP + TN + FP + FN}$$

(3)

2) **Precision**: The fraction of predicted exoplanet detections that are correct.

$$\text{Precision} = \frac{TP}{TP + FP}$$

(4)

3) **Recall (Sensitivity or True Positive Rate):** The proportion of actual exoplanets that are correctly identified by the model.

$$\text{Recall} = \frac{TP}{TP + FN}$$

(5)

4) **F1 Score:** The harmonic mean of precision and recall, balancing both metrics.

$$F1 = 2 \times \frac{\text{Precision} \times \text{Recall}}{\text{Precision} + \text{Recall}} \quad (6)$$

5) **Confusion Matrix:** A table that summarizes the performance of the model by showing TP, FP, TN, and FN counts

$$\begin{bmatrix} TP & FP \\ FN & TN \end{bmatrix} \quad (7)$$

## VI. RESULTS

In this study we evaluated several machine learning models aimed at detecting exoplanets concerning two important performance metrics, accuracy and F1 score. Although accuracy describes the ratio between correct predictions and total predictions, making it a good overall gauge of model performance, we focused especially on the F1 score as a better metric for evaluation because exoplanet data more often provides a class imbalance where verified planets are few in relation to observations.

The F1 score gives a more balanced performance measure by balancing precision and recall. Precision is a measure of the model's ability to avoid false positive errors (non-planets that were mistaken for planets) and recall gives a measure of its ability to find true planetary signals (the true planet rate that it has identified correctly). By providing these two complementary measures in one holistic score, the F1 score provides a more comprehensive, nuanced measure of model performance when working with the heavily skewed datasets that typify astronomical measurements of exoplanetary systems.

The CNN model had strong classification power with a 99.5 percent accuracy level, which indicates its excellent ability to correctly classify instances in our exoplanet detection task. The high accuracy is consistent with the CNN's established strength in recognizing and extracting complex patterns from the stellar light curve data. However, upon examining the model's F1 score of 0.727, we observed a large difference between this balanced metric and the overall accuracy measurement.

This performance gap reveals major subtleties in the performance of the model, i.e., challenges in achieving maximum balance between precision and recall in the detection of exoplanets. While very close to achieving perfect accuracy, the CNN struggled with consistency in detecting actual exoplanetary signals as opposed to false positives. The difference between these metrics clearly shows that while its overall performance is outstanding, the model must be optimized to become more reliable. Optimizations can be through architectural changes, improved feature selection, or more sophisticated data augmentation methods specifically designed to minimize the rate of false positives without impairing the model's ability to detect actual planetary transits in stellar observation data.

For comparison, the Random Forest model, which uses an ensemble of decision trees, was 97.8 percent accurate, slightly less than that of the Convolutional Neural Network. Yet its F1 score was 0.790, greater than that of the CNN, which indicates a more balanced precision-recall. The greater F1 score means the Random Forest model was more accurate in labeling exoplanets and non-exoplanets and therefore a more balanced option for this specific dataset where precision is as valuable as recall.

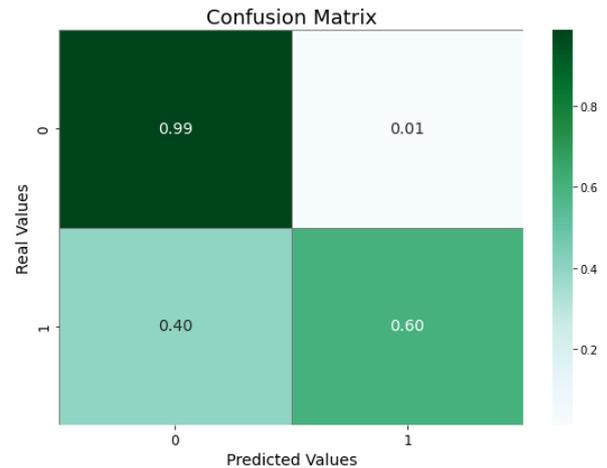

Fig. 8. Confusion Matrix

The LightGBM model, which is another popular classification algorithm, also recorded 96.5% accuracy but with a lower F1 score of 0.675. This shows that although the LightGBM was mostly accurate in classifying the data, it performed poorly in the precision and recall balance, especially in separating the actual exoplanets from the false positives. This is a typical problem with imbalanced datasets, where a small number of false positives would have a massive effect on the F1 score.

Finally, the Logistic Regression model was 95.2% accurate and had an F1 score of 0.640, the least balanced of all the models employed. Although Logistic Regression is a good baseline model to employ in classification problems, the fact that this research had a relatively lower F1 score from it means that perhaps it is not the most suitable model to be employed in exoplanet detection, particularly in dealing with unbalanced data.

Finally, although the CNN model was most accurate, the superior F1 score of the Random Forest model would indicate that it is more balanced between precision and recall. Although the LightGBM and Logistic Regression models were quite accurate, their lower F1 scores would indicate more challenges in balancing false positives and false negatives to identify exoplanets. It is in this context that the importance of taking accuracy and F1 score into consideration is highlighted, particularly in situations where class imbalances have a significant effect on model performance.

| Model | Accuracy(%) | F1 Score |
|---|---|---|
| Convolutional Neural Network (CNN) | 99.5 | 0.727 |
| Random Forest | 97.8 | 0.790 |
| LightGBM | 96.5 | 0.675 |
| Logistic Regression | 96.2 | 0.640 |
| Support Vector Classifier [2] | 98.9 | 0.760 |

Table. 1. Experimentation Results

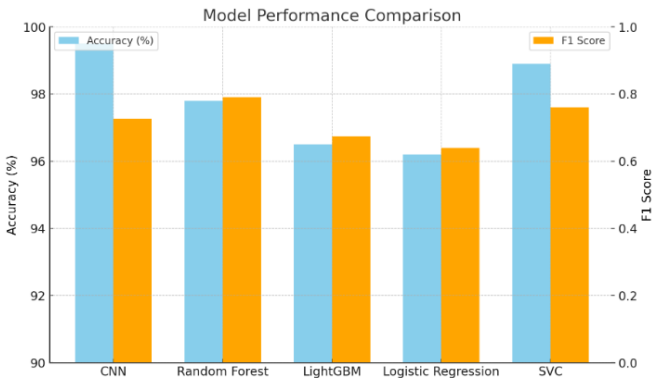

Fig. 9. Model Performance Comparison

## VII. CONCLUSION

Herein, we adopted a twofold approach marrying machine learning and deep learning architectures to overcome the drawbacks of exoplanet detection. The hybrid approach was precisely crafted to utilize the complementary merits of various algorithmic paradigms in creating a stronger detection framework optimizing prediction performance and robustness. Using both traditional machine learning and more modern deep learning models, we gained a better assessment of model performance in managing the complex, high-dimensional, and highly unbalanced data typical of astronomical data, where validated planetary signals constitute a minority of the overall data.

One of the key focuses of this research, aside from our machine learning and deep learning detection algorithms, is the derivation of planetary properties from light data observations for exoplanet characterization. We mostly analyze light curves, which indicate stellar brightness variations as a function of time. When an exoplanet transits its parent star, it creates a distinctive brightness dip that our advanced algorithms recognize and analyze to deduce planetary properties and confirm exoplanets orbiting distant stars.

In this particular study, the ML and DL models application in light curve data analysis offers higher detection efficiency and also allows us to extract the key properties of the planets. From these properties such as size of the planet, orbital determinants, atmospheric composition, we can extract key knowledge to assess the actual nature of these alien planets and their ability to nurture life. Future works would want to minimize the uncertainties of such measurements through additional sources of data like radial velocity curves and strengthen the ML and DL models to account for the adverse effects of noisy and varying light data. The extraction of planetary properties from light curve data goes a long way to deepening knowledge in exoplanetary studies. With the advent of newer ML and DL algorithms, we can actually deduce more solid estimates for several planetary properties from observations.

Future investigations can try toward refining the accuracy of such estimates through more input data, such as from radial velocity measurements, as well as advancing the ML and DL models further toward handling noisy and complex light signals. Geared at enhancing not just the precision in exoplanet detection but also the comprehension toward such worlds and their potential for life.